\documentclass[aps,pre,twocolumn,groupedaddress]{revtex4-1}
\usepackage{graphicx}
\usepackage{amssymb}
\usepackage{amsmath}
\usepackage{bm}
\usepackage[caption=false]{subfig}
\begin{document}
\title{Topological phase transitions induced by the variation of exchange couplings in graphene}

\author{Jihyeon \surname{Park}}
\affiliation{Department of Physics, Ewha Womans University, Seoul 03760, Korea}
\author{Gun Sang \surname{Jeon}}
\email[gsjeon@ewha.ac.kr]{}
\affiliation{Department of Physics, Ewha Womans University, Seoul 03760, Korea}

\begin{abstract}
We consider a modified graphene model under exchange couplings.
Various quantum anomalous phases are known to emerge under uniform or staggered
exchange couplings.
We introduce the twist between the orientations of two sublattice exchange couplings, which is
useful for examining how such topologically nontrivial phases under different
types of exchange couplings  are connected to one another.
The phase diagrams constructed by the variation of exchange coupling strengths and
twist angles exhibit rich structures of successive topological transitions. 
We analyze the emergence of peculiar phases in terms of the evolution of the
energy dispersions.
Perturbation schemes applied to the energy levels turn out to reproduce well 
phase boundary lines up to moderate values of the twist angle.
We also discover two close topological transitions under uniform exchange
couplings,
which is attributed to the interplay of the trigonal-warping deformation due to
Rashba spin-orbit coupling and the staggered sublattice potential.
Finally the implications of Berry curvature structure and topological
excitations in real and pseudo spin textures are discussed.
\end{abstract}

\flushbottom
\maketitle
\thispagestyle{empty}

\section{Introduction}
Quantum anomalous Hall effect is a variation of quantum Hall effect which occurs
with spontaneously broken time-reversal symmetry in the absence of external
magnetic field~\cite{Chang2023,Liu2016,Liu2008a,Onoda2003,Wu2008}. It is
distinguished from quantum Hall effect which requires strong external magnetic
field and quantum spin Hall effect which appears in the presence of
time-reversal
symmetry~\cite{Klitzing1980,Thouless1982,Qiao2010,Sheng2006,Qiao2011a}. Quantum
anomalous Hall effect makes Chern insulators have dissipationless chiral edge
states and insulating bulk states, which is characterized by Chern
number~\cite{Chang2023}. Chern number $C$ is physically related to Hall
conductivity $\sigma_{xy}$ via 
$\sigma_{xy}=C\frac{e^{2}}{h}$~\cite{Bernevig2013,Fradkin2013,Vanderbilt2018}. 

Many candidates have been suggested as materials to exhibit quantum anomalous
Hall effect and some of them were successful~\cite{Chang2023}. Since quantum
anomalous Hall effect requires band inversion and time-reversal symmetry
breaking, it can be naturally considered to catalyze magnetism in topological
materials to realize it. 
Magnetically doped topological
insulators such as Cr-doped $\mathrm{(Bi,Sb)}_{2}\mathrm{Te}_{3}$ films first
showed quantum anomalous Hall effect~\cite{Chang2013}. The observation
of quantum anomalous Hall effect in intrinsic magnetic topological insulator
such as $\mathrm{MnBi_{2}Te_{4}}$ flakes was also reported~\cite{Deng2020}. 
Recently, moir$\acute{\mathrm{e}}$ materials are expected
to host quantum anomalous Hall effect due to their strong correlations to break
time-reversal symmetry and realized in the heterostructure of
hexagonal boron nitride~\cite{Serlin2020}.  

Several pioneering studies motivated
extensive theoretical studies on the compounds with honeycomb-type lattice
structure and strong spin-orbit
coupling~\cite{Haldane1988a,Kane2005a}. 
In this context graphene was
proposed to exhibit quantum anomalous Hall effect in the presence of Rashba spin
orbit coupling and exchange coupling~\cite{Qiao2010,Qiao2012}. This model shows
gap opening and nontrivial Berry curvature in the vicinity of $K$ and
$K^{\prime}$ in the hexagonal Brillouin zone~\cite{Qiao2010}. The Berry
curvature is integrated to produce a nontrivial Chern number 
in the system, which
characterizes quantum anomalous Hall effect. 
Such theoretical models are expected
to be realized by the addition of  transition-metal atoms on top of
graphene~\cite{Qiao2010, Chang2023}; it has not been observed yet in real
materials. 
However, germanene which also has honeycomb lattice was reported recently to
host quantum spin Hall effect~\cite{Bampoulis2023}. 

The graphene model with quantum anomalous Hall effect can be extended with the
additional
intrinsic spin orbit coupling and staggered sublattice
potential~\cite{Qiao2012,Hogl2020}. 
While intrinsic spin orbit coupling in
pristine graphene is weak, proximity spin orbit coupling in
graphene induced by transition-metal dichalcogenides can be intensified in meV
scale. Besides, the proximity spin orbit coupling acquires staggered form on
sublattices A and B~\cite{Hogl2020}. Meanwhile, exchange coupling can be
either uniform or staggered depending on the magnetism of
substrates~\cite{Hogl2020}. 
Based on these facts, topological phases
under uniform and staggered regime of
intrinsic spin orbit coupling and exchange coupling were
investigated~\cite{Hogl2020}. 
As a result, a variety of interesting  quantum anomalous
Hall phases were  predicted such as those with Chern number two 
in uniform intrinsic spin orbit coupling and
uniform exchange coupling, and those with Chern number one in uniform
intrinsic spin orbit coupling and staggered exchange coupling~\cite{Hogl2020}. 
One may lead to questions as to 
whether such nontrivial phases are connected continuously to one another
and how the phases evolves during the path,
which is one of the main motivations of our study.

In this paper, we investigate the topological phase transition of the modified graphene
model with quantum anomalous Hall effect by varying the relative orientation of
exchange couplings of two sublattices. 
Rich phase diagrams are obtained by the numerical diagonalization.
Topologically nontrivial phases are characterized by Chern numbers, and
the change in Chern numbers  are discussed in terms of the
touching of valence and conduction bands.
The topological phase transitions for small twist angles are explained 
quantitatively by the perturbation theory.
Two successive transitions as well as distorted trigonal-warping deformation are
also found to take place for small twist angles.
We scrutinize the nature of topological phases  in terms of the distribution of
Berry curvature for valence bands and topological objects in real and pseudo
spin textures. 

\section{Model}
We consider the half-filled proximity-modified graphene model described by the Hamiltonian
\begin{equation}
		H=H_{0}+H_{R}+H_{S}+H_{I}+H_{E}\label{eq_H_qahe}
\end{equation}
with
\begin{eqnarray}
		H_{0}&=&-t\sum_{\langle i,j\rangle,\alpha}c^{\dagger}_{i\alpha}c_{j\alpha},
		\\
	  	H_{R}&=&i \lambda_{R}\sum_{\langle
		  i,j\rangle,\alpha,\beta}c^{\dagger}_{i\alpha}c_{j\beta}[\small(\hat{\bm{\sigma}}\times\hat{\bm{d}}_{ij})_{z}]_{\alpha\beta},
		\\
			  H_{S}&=&\Delta\sum_{i,\alpha}\xi_{i}c^{\dagger}_{i\alpha}c_{i\alpha},
		\\
			  H_{I}&=&\frac{i\lambda_{I}}{3\sqrt{3}}\sum_{\langle\langle
			  i,j\rangle\rangle,\alpha,\beta}\nu_{ij}c^{\dagger}_{i\alpha}c_{j\beta}[\hat{\sigma}_{z}] _{\alpha\beta},
		\\
			  H_{E}&=&\lambda_{E}\sum_{i,\alpha,\beta}c^{\dagger}_{i\alpha}c_{i\beta}[
				  \hat{\bm{m}}_{i}\cdot\hat{\bm{\sigma}}]_{\alpha\beta}.
\end{eqnarray}
Here, $c_{i\alpha}^{\dagger}(c_{i\alpha})$ is the creation(annihilation)
operator of an electron with spin $\alpha$ at site $i$ on the honeycomb lattice. 
$H_{0}$ describes the
hopping between the nearest neighbor sites and the summation runs over all the
nearest neighbor pairs $\langle i,j\rangle$. 
$H_{R}$ represents the Rashba spin orbit coupling of strength $\lambda_R$ where $\hat{\bm{\sigma}}$ is
the vector whose components are Pauli matrices and $\hat{\bm{d}}_{ij}$ is the
unit vector of the path from site $j$ to $i$. 
$H_{S}$ denotes the staggered
sublattice potential of strength $\Delta$ with 
\begin{equation}
\xi_i=
		  \begin{cases}
					 +1 & \hbox{for } i \in \hbox{A},
					 \\
					 -1 & \hbox{for } i \in \hbox{B}.
		  \end{cases}
\end{equation}
$H_{I}$
indicates the intrinsic spin orbit coupling between next nearest neighbors
with the summation over all the pairs $\langle\langle i,j\rangle\rangle$ and
$\nu_{ij}=\pm 1$ when the
path from site $j$ to $i$ bends counterclockwise/clockwise. 
$H_{E}$ describes exchange couplings of strength $\lambda_E$ in the direction
$\hat{\bm{m}}_{i}\equiv(\cos{\phi_{i}}\sin{\theta_{i}},\sin{\phi_{i}}\sin{\theta_{i}},\cos{\theta_{i}})$
 at site $i$. 

 In this work we will employ the twisted exchange couplings where the exchange
 couplings are oriented in $z$ direction at sublattice A $(\theta_{i}=0,\phi_{i}=0)$
 and it is twisted by
 the angle $\theta_T$ about the $y$ direction at sublattice B
 $(\theta_{i}=\theta_{T},\phi_{i}=0)$; 
 this corresponds to
\begin{equation}
		  \hat{\bm{m}}_{i}=
	\begin{cases}
			  (0,0,1) & \hbox{for } i\in \hbox{A},\\
			  (\sin{\theta_{T}},0,\cos{\theta_{T}}) & \hbox{for } i\in \hbox{B}.
	\end{cases}
		  \label{eq_exchange}
\end{equation}
The uniform and the staggered exchange couplings
correspond to the twisted exchange couplings with 
$\theta_{T}=0$ and $\theta_{T}=\pi$, respectively. 
By the continuous variation of the twist angle $\theta_T$
we can conveniently examine how the topological phases evolve between 
the uniform and the staggered exchange couplings.

Henceforth we will focus on two values of the uniform intrinsic spin orbit couplings
$\lambda_I=-0.05t$ and $0.05t$ for sublattice potential $\Delta=0.1t$ and Rashba
spin-orbit coupling $\lambda_R=0.05t$.
In the earlier work~\cite{Hogl2020}
the uniform exchange coupling was shown to result in the same topological
transitions for both cases.
On the other hand, in the presence of the staggered exchange couplings
the resulting intermediate topological phases display different topological
invariants. 
We examine the topological phase transitions by varying the twist angle
$\theta_T$ with particular attention to the two cases, which will help us to 
understand the underlying physical implications in a variety of topological
phase transitions depending on the patterns of exchange couplings.
Throughout the paper, we measure all
the energy scales in units of the hopping strength $t$ between nearest neighbors
and the length scales in units of next-nearest-neighbor spacing $a$.

\section{Results}
\subsection{Phase Diagram}

\begin{figure}
	\begin{center}
			  \subfloat[]{
				\includegraphics[width=7.5cm]{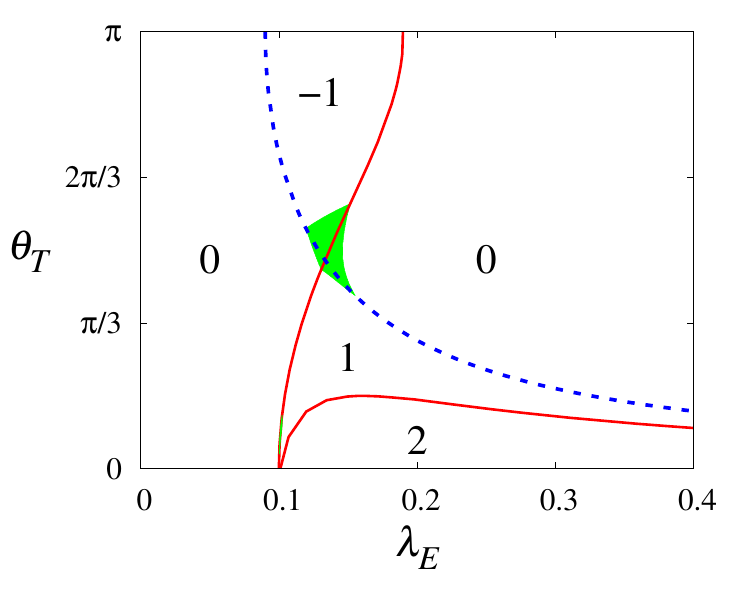}
			}

				  \subfloat[]{
				\includegraphics[width=7.5cm]{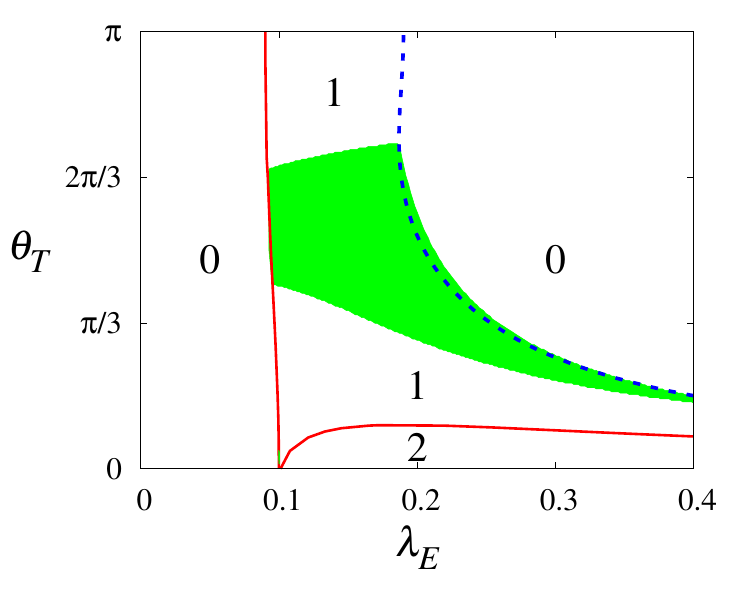}
			}
	\end{center}
	\caption{Phase diagrams of the proximity-modified graphene with twisted exchange 
			  couplings of strength $\lambda_E$ and twist angle $\theta_{T}$ 
			  for $\Delta=0.1$, $\lambda_R=0.05$, 
			  and (a) $\lambda_I=-0.05$; (b) $\lambda_I=0.05$. 
			  Red solid and blue dashed lines represent phase transition lines
			  induced by band crossing near $K$ and $K^{\prime}$, respectively. 
			  The area marked in green denotes a metallic region.
			  The numbers displayed indicate the Chern numbers of the
			  corresponding topological phases.
			  }
	\label{pd}
\end{figure}

The topological phases are characterized by Chern number defined by
\begin{align}
	C =\frac{1}{2\pi}\sum_{n}\int_{\textrm{BZ}}d^{2}k \ \ \Omega_{n}(\bm{k}), 
	\label{eq_C}
\end{align}
where 
the summation of $n$ runs over all the filled valence bands
and $\Omega_{n}(\bm{k})$ is 
Berry curvature of the $n$th valence band at
momentum $\bm{k}$, defined by
\begin{equation}
	\Omega_{n}(\bm{k}) {=}
		-2\sum_{n^{\prime}\neq n}
		\frac{
			\textrm{Im}\langle\psi_{n,\bm{k}}|
			\partial_{k_{x}} H_{\bm{k}}
			|\psi_{n^{\prime},\bm{k}}\rangle
			\langle\psi_{n^{\prime},\bm{k}}|\partial_{k_{y}}
			H_{\bm{k}}|\psi_{n,\bm{k}}\rangle
		}{
			(E_{n^{\prime},\bm{k}} - E_{n,\bm{k}})^{2}
		},
	\label{eq_BC}
\end{equation}
with the eigenenergy $E_{n,\bm{k}}$ and the eigenfunction $\psi_{n,\bm{k}}$.
By the exact diagonalization method, we obtain the eigenvalues and eigenvectors
of the Fourier transformed Hamiltonian $H_{\bm{k}}$. 
Numerical integration of Berry curvature is performed over the Brillouin zone,
which yields the Chern number of the phase. 

Phase diagrams are constructed by the resulting Chern numbers for various
exchange coupling strengths $\lambda_E$ and twist angles $\theta_T$.
In Fig. \ref{pd} we plot two phase diagrams for $\lambda_I=\pm0.05$ as mentioned
in the previous section.
The two systems have common behaviors in the topological phase transitions in
the limits of small and large $\lambda_E$.
For small $\lambda_E$ the system generally displays zero Chern number. 
On the other hand, for large $\lambda_E$, 
the system exhibits $C=2$ in the presence of uniform exchange coupling
($\theta_T=0$).
As $\theta_T$ increases, the system undergoes two successive topological transitions and
Chern number reduces by one at each transition.
Thus, for staggered exchange coupling $(\theta_T=\pi)$, the resulting phase is
topologically trivial in both limits.

In the intermediate region of exchange coupling strength $\lambda_E$
the topological characters of the two systems with $\lambda_I =\pm 0.05$ are
very different.
The system with $\lambda_I=-0.05$ exhibits three successive transitions from
$C=2$ with the increase of $\theta_T$ and accordingly we obtain $C=-1$ for 
$\theta_T=\pi$.
For $\lambda_I=0.05$, in contrast, only a single topological transition occurs
with increasing $\theta_T$ 
and the phase with $C=1$ persists up to $\theta_T=\pi$ without further
transitions. 

At phase boundaries where Chern number changes by one the lower conduction band
and the upper valence band touches at one point $\bm{k}_0$.
One can find that we find two phase boundaries where $\bm{k}_0$ is near $K$
point (displayed in red solid lines) and one with $\bm{k}_0$ being near $K^\prime$
(displayed in blue dashed lines).
It is of interest to note that $\bm{k}_0$ is located exactly at the symmetric
point ($K$) only on the left red solid line.
On the other two phase boundaries, $\bm{k}_0$ changes with $\lambda_E$ although
$\bm{k}_0$ is close to $K$ or $K'$ in the region displayed in Fig.~\ref{pd}.
We have obtained the precise positions of 
the phase boundaries by numerically identifying the value of 
$\lambda_{E}^{c}(\theta_T)$ for which the conduction and the valence bands 
touch each other and constructed the phase diagram in Fig.~\ref{pd}.

The analysis of the variation of the phase boundaries with $\theta_T$ reveals
the the origin of the different behavior in the intermediate regions of
$\lambda_E$.
As illustrated in Fig.~\ref{pd}, we have one red($K$) and one blue($K'$) boundary lines
which traverse the whole range of $\theta_T$ between the uniform and the staggered
exchange couplings.
For $\lambda_I=0.05$ both $\lambda_E^c(K)$ and $\lambda_E^c(K')$ decreases with the increase of $\theta_T$ and do not cross each other.
Thus the phase with $C=1$ for small $\theta_T$ extends continuously to
$\theta_T=\pi$.
For $\lambda_I=-0.05$, on the other hand, $\lambda_E^c(K)$ increases with
$\theta_T$ and the resulting phase boundary crosses the $K'$ boundary line.
The crossing point results in two more successive topological transitions
and the system exhibits $C=-1$ at $\theta_T=\pi$

\begin{figure}
 	\begin{center}
 		\includegraphics[width=8cm]{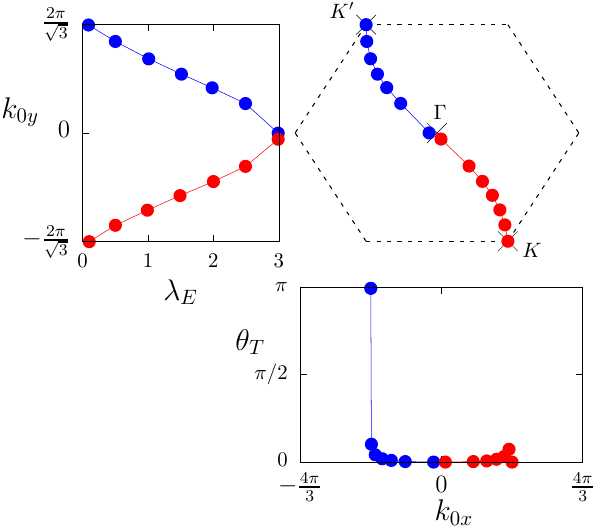}
 	\end{center}
 	\caption{
			  The band-touching momentum $\bm{k}_0$ of the proximity-modified graphene model with twisted
			exchange couplings on the phase boundaries as a function of $\lambda_E$ and
			$\theta_T$ 
			  for $\Delta=0.1$, $\lambda_R=0.05$, 
			and $\lambda_{I}=-0.05$. 
			A dashed hexagon indicates the boundary of the Brillouin zone
			and the crosses($\times$) indicate the location of symmetric points, $K$,
			$K^{\prime}$, and $\Gamma$ in the Brillouin zone. } 
			\label{locus}
			
 \end{figure}
 
Another interesting feature in the phase diagram is the existence of metallic
regions for intermediate $\theta_T$.
The metallic phase shows up when the minimum of the conduction band is lower
than the maximum of the valence band, which yields partial filling in the
conduction band without the overlap of valence and conduction bands.
For $\lambda_I=-0.05$ the metallic region is located around the crossing point
of two traversing $K$ and $K'$ phase boundaries.
Such a metallic region also appears for $\lambda_I=0.05$ in the middle of the
region with $C=1$, and it separates $C=1$ phase for uniform exchange coupling
from that for staggered exchange coupling.

\begin{figure}
	\begin{center}
			  \subfloat[]{\includegraphics[width=7.5cm]{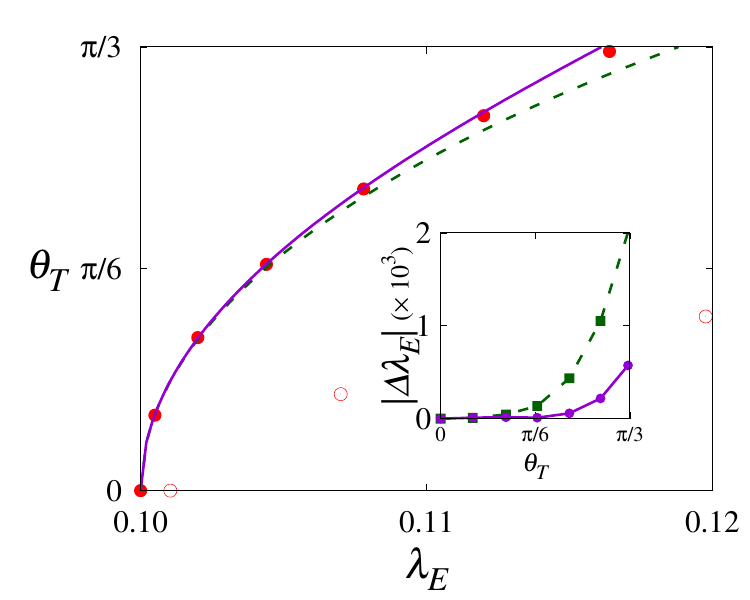}}

			  \subfloat[]{\includegraphics[width=7.5cm]{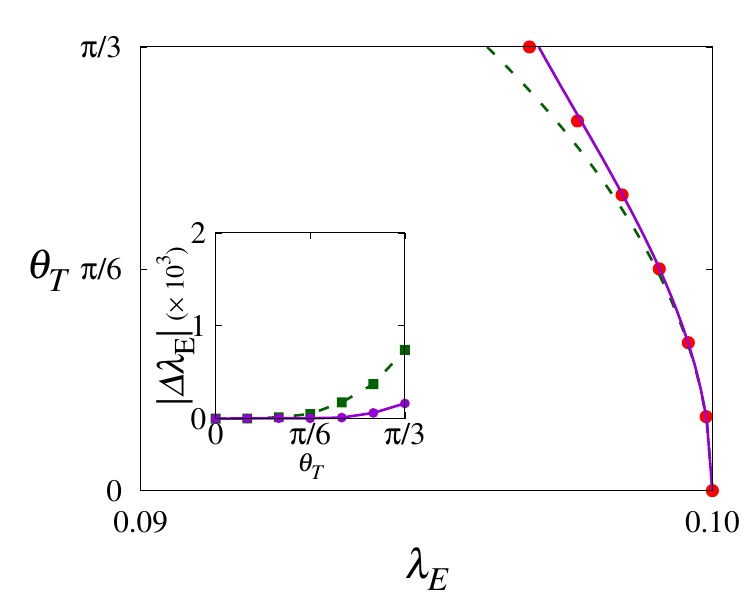}}
	\end{center}
	\caption{Comparison of phase boundaries 
			from exact diagonalization and perturbative calculations
		  in the proximity-modified graphene model with twisted exchange couplings
for $\Delta=0.1$, $\lambda_R=0.05$ and (a) $\lambda_I=-0.05$; (b)
$\lambda_I=0.05$.
		  The phase boundaries estimated from exact diagonalization are marked by
		  red circles, and green dashed and purple solid lines denote
		  those obtained from the second-order and the fourth-order perturbation
		  theory in the twist angle $\theta_T$,  respectively.
		  The insets show the absolute differences between the numerical and the
		  perturbative values of $\lambda_E^c$ as a function of $\theta_T$.
		  }  
		  \label{pd_anal}
\end{figure}

Figure~\ref{locus} displays how the band-touching momentum $\bm{k}_0$ changes as
the system parameter varies.
For small $\lambda_E$ two bands touch around $K$ point on the ``red'' boundary
and around $K'$ point on the ``blue'' boundary.
As $\lambda_E$ increases, both $|k_{0x}|$ and $|k_{0y}|$ reduce and $\bm{k_0}$
monotonically approaches $\Gamma$ point.
Although $K'$ boundary line starts from $\theta_T=\pi$, we can find that the
twist angle $\theta_T$ drastically decreases with the increase of $\lambda_E$.
As demonstrated in Fig.~\ref{locus} the band-touching momentum is close to
$\Gamma$ for $\lambda_E \gtrsim 3$.

 		  \subsection{Perturbation Theory}
In this section, we apply the perturbation theory to obtain 
the phase boundary which is determined by the band-touching at $K$ point.
The characteristic equation of the Hamiltonian at $K$ is given by
\begin{align}
	\begin{split}
	\big(&\Delta-\lambda_{I}-\lambda_{E}-E\big)\times 
	\\
	\big[ &\big(\Delta+\lambda_{I}+\lambda_{E}-E\big)\times
	\\
	&\big(-\Delta-\lambda_{I}+\lambda_{E}\cos{\theta_{T}}-E\big)\times
	\\
	&\big(-\Delta+\lambda_{I}-\lambda_{E}\cos{\theta_{T}}-E\big)
	\\
	&-9\lambda_R^{2}\big(-\Delta-\lambda_{I}+\lambda_{E}\cos{\theta_{T}}-E\big)
	\\
	&-\lambda_{E}^{2}\sin^{2}{\theta_{T}}\big(\Delta+\lambda_{I}+\lambda_{E}-E\big)\big]=0,\label{eq_ch_K}
	\end{split}
\end{align}
where $E$ is an energy eigenvalue. 

For $\theta_T=0$, 
four energy levels are given by
\begin{equation}
	\begin{split}
		E_1^{(0)}=&\lambda_{I}-\sqrt{(\Delta+\lambda_{E})^{2}+9\lambda_R^{2}},\\
		E_2^{(0)}=&-\lambda_{I}-\Delta+\lambda_{E},\\
		E_3^{(0)}=&-\lambda_{I}+\Delta-\lambda_{E},\\
		E_4^{(0)}=&\lambda_{I}+\sqrt{(\Delta+\lambda_{E})^{2}+9\lambda_R^{2}},\\
	\end{split}\label{eq_ev_K_0}
\end{equation}
and the topological transition occurs at $\lambda_E=\Delta$ by the band-crossing
of $E_2^{(0)}$ and $E_3^{(0)}$.

We apply the perturbation theory by trying the power-series solution of the
energy eigenvalues
\begin{equation}
	E_i = E_i^{(0)} + \sum_{n=1}^\infty c_i^{(n)} \theta_T^n \quad \quad
(i=1,2,3,4).
\end{equation}
Since the characteristic equation is an even function of $\theta_T$, $c_i^{(n)}=0$ for odd $n$.
From the overall factor in Eq.~(\ref{eq_ch_K})
we can also find that $E_3=E_3^{(0)}$ is independent of $\theta_T$.

By inserting ${E_2}$ to Eq.~(\ref{eq_ch_K}) and expanding it to the fourth order
in $\theta_{T}$, we find the first two nonvanishing coefficients
\begin{equation}
	\begin{split}
		c_2^{(2)}=&-\frac{\lambda_{E}}{2}-\frac{2\lambda_{E}^{2}\big(\Delta+\lambda_{I}\big)}{4\big(\lambda_{I}-\lambda_{E}\big)\big(\Delta+\lambda_{I}\big)-9\lambda_R^{2}},
\\
	c_2^{(4)}=&\frac{\lambda_{E}}{24}+
\frac{1
}{
	4(\lambda_{I}-\lambda_{E})(\Delta+\lambda_{I})-9\lambda_R^{2}
} \times
\\
&\Big[
\lambda_{E}^{2}(\Delta+\lambda_{I})/6+c_{2}^{(2)}\lambda_{E}\lambda_{I}
\\
& \quad +2\left(c_{2}^{(2)}\right)^{2}(2\lambda_{I}-\lambda_{E}+\Delta)
\Big].
\end{split}
\end{equation}  

Figure~\ref{pd_anal} displays the results from the perturbation calculation of the
second and the fourth order in $\theta_{T}$ for $\lambda_I=\pm0.05$. 
We can observe that 
second-order perturbation results reproduce the phase boundaries 
well at least up to $\theta_{T}=\pi/6$. 
The fourth-order results show better agreement for higher
$\theta_{T}$ than the second-order ones. 
It is interesting that this approach identifies only
one of two phase boundaries which split near 
$(\lambda_{E}=\Delta$ and $\theta_{T}=0)$. 
The reason is that the phase boundary denoted by open circles is caused by the
band-touching which does not occur exactly at $K$ point.
We will examine the peculiar features of this phase boundary in the next
section.

		  \subsection{Fine structures near the transition under
uniform exchange coupling}

\begin{figure}
	  \subfloat[]{\includegraphics[width=7.5cm]{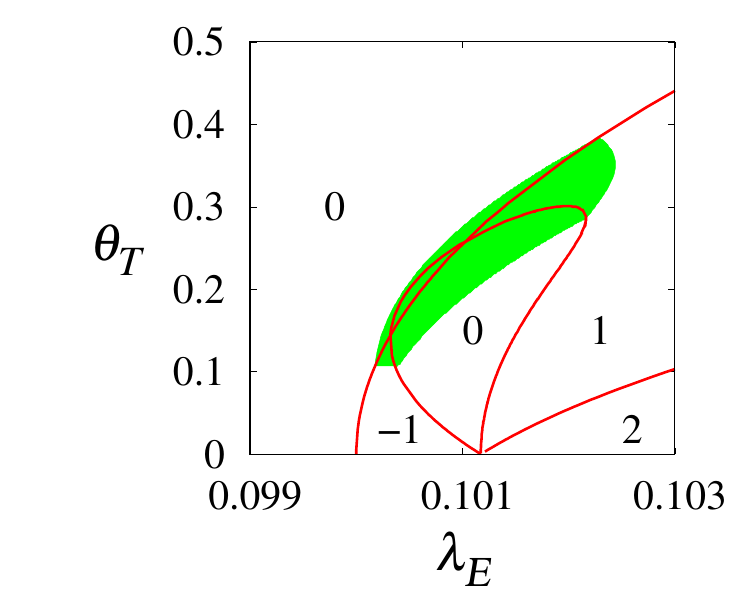}}

	  \subfloat[]{\includegraphics[width=7.5cm]{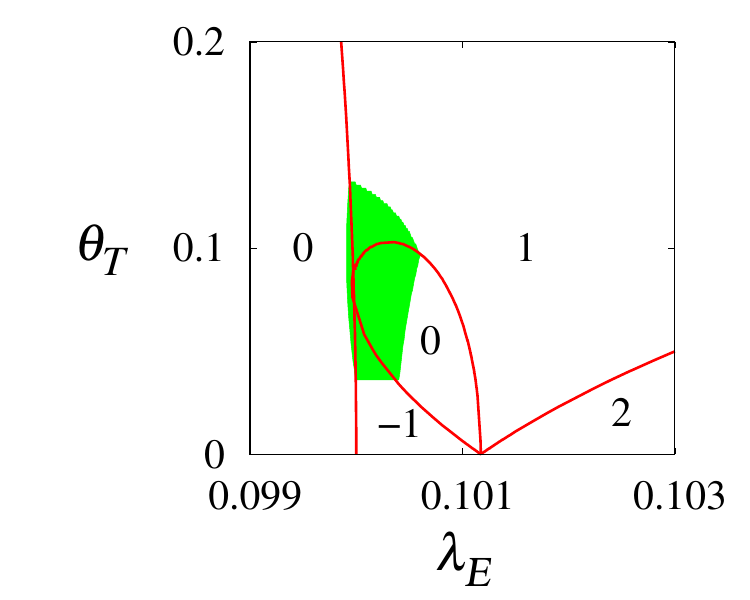}}
	  \caption{Phase diagram magnified in the vicinity of the transition point at
		  $\theta_T=0$.
		  The number displayed is the Chern number of the corresponding phase.
			The areas marked in green denote metallic regions.
	  }
	  \label{PD2}
\end{figure}

\begin{figure}
	  \subfloat[]{\includegraphics[width=4.3cm]{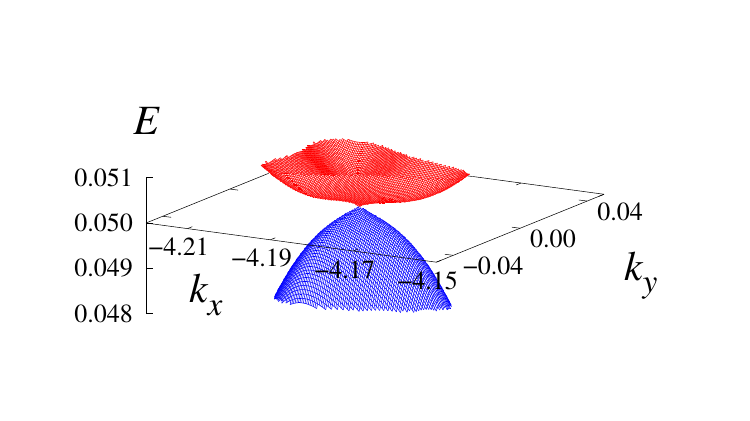}}
	  \subfloat[]{\includegraphics[width=4.3cm]{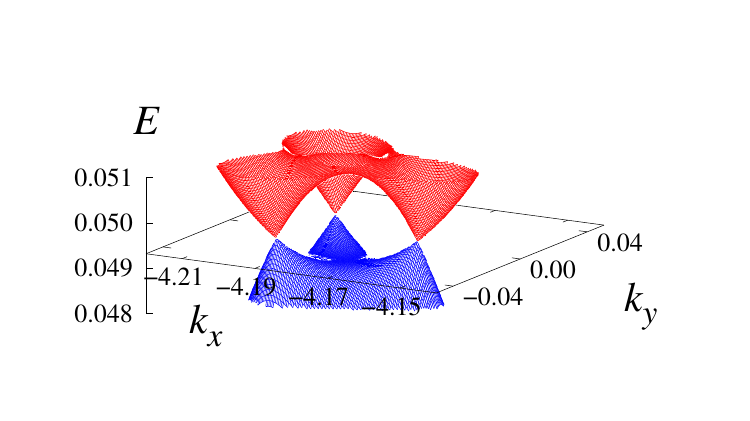}}

	  \subfloat[]{\includegraphics[width=4.3cm]{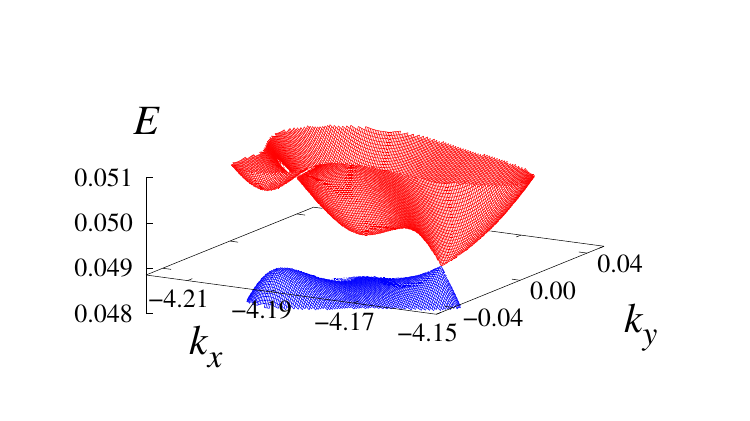}}
	  \subfloat[]{\includegraphics[width=4.3cm]{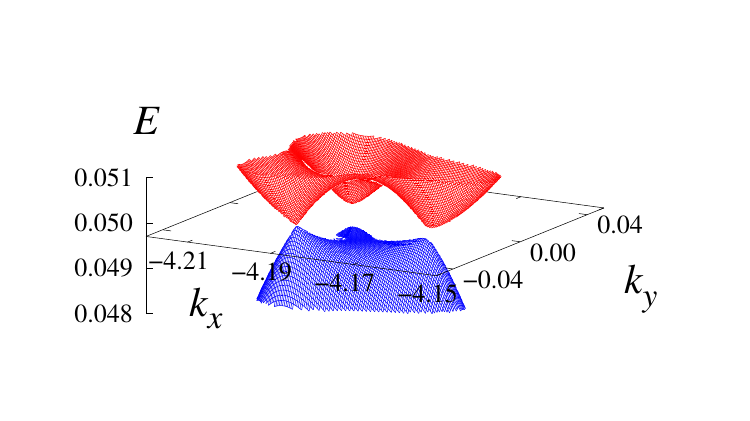}}
	  \caption{
				 Energy dispersions of the proximity-modified graphene model with
twisted exchange couplings for $\Delta=0.1$, $\lambda_R=0.05$,
				 (a) $\lambda_E=0.1$, and $\theta_T=0$; 
				 (b) $\lambda_E=0.1012$, and $\theta_T=0$; 
				 (c) $\lambda_E=0.101655$, and $\theta_T=0.1775$; 
				 (d) $\lambda_E=0.10074$, and $\theta_T=0.04125$. 
	  }
	  \label{tri_ED}
\end{figure}

\begin{figure}
		  \includegraphics[width=8.5cm]{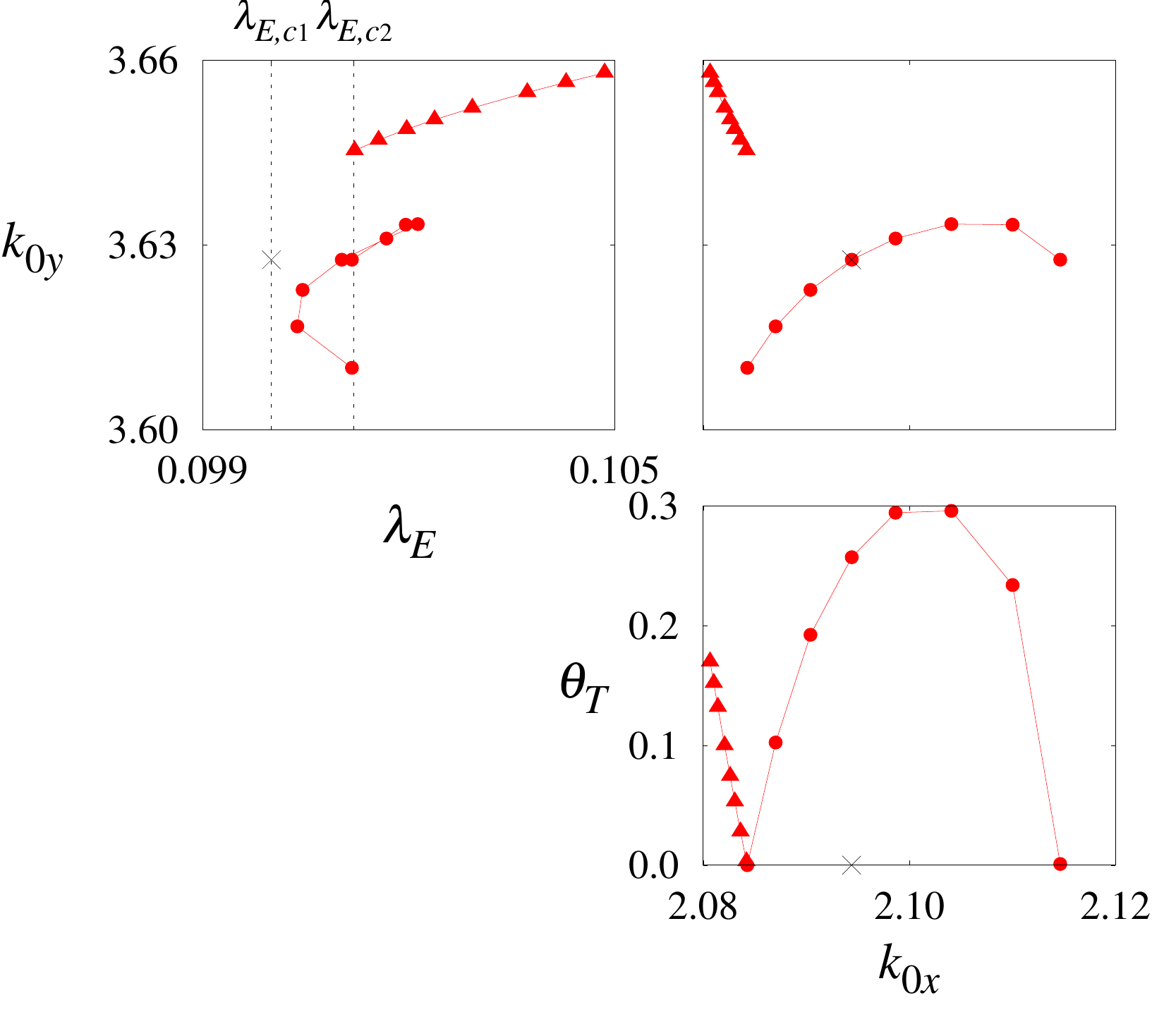}
		  \caption{
					 The band-touching momentum $\bm{k}_0$  of the proximity-modified graphene model with twisted
					 exchange couplings on the phase boundaries as a function of
					 $\lambda_E$ and $\theta_T$ 
			  for $\Delta=0.1$, $\lambda_R=0.05$, 
			and $\lambda_{I}=-0.05$. 
The dotted lines are the two transition points $\lambda_{E,c1}$ and
$\lambda_{E,c2}$ at $\theta_T=0$, and the cross($\times$) indicates $K$ point.
 		}
\label{GPlocus2}
\end{figure}

Figure~\ref{PD2} presents the phase diagram magnified in the vicinity of the
topological transition at $\theta_T=0$.
It is remarkable that for the uniform exchange coupling ($\theta_T=0$) 
the system does not exhibit a direct transition from
a topologically trivial phase ($C=0$) for small $\lambda_E$ 
to a topological phase ($C=2$) for large $\lambda_E$.
As $\lambda_E$ increases, the system undergoes a transition to a
topological phase with $C=-1$ at $\lambda_{E,c1}=0.1$, and successively 
to a second topological phase with $C=2$ at $\lambda_{E,c2}=0.1012(1)$.

 \begin{figure*}
	\includegraphics[width=8.5cm]{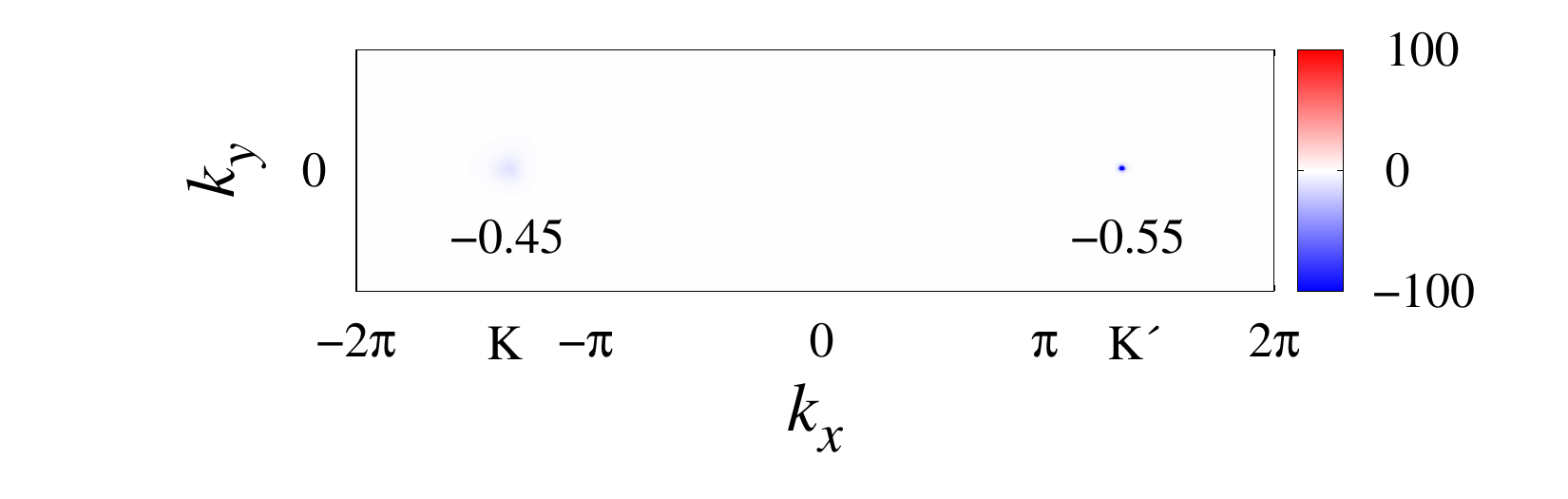}
	\includegraphics[width=8.5cm]{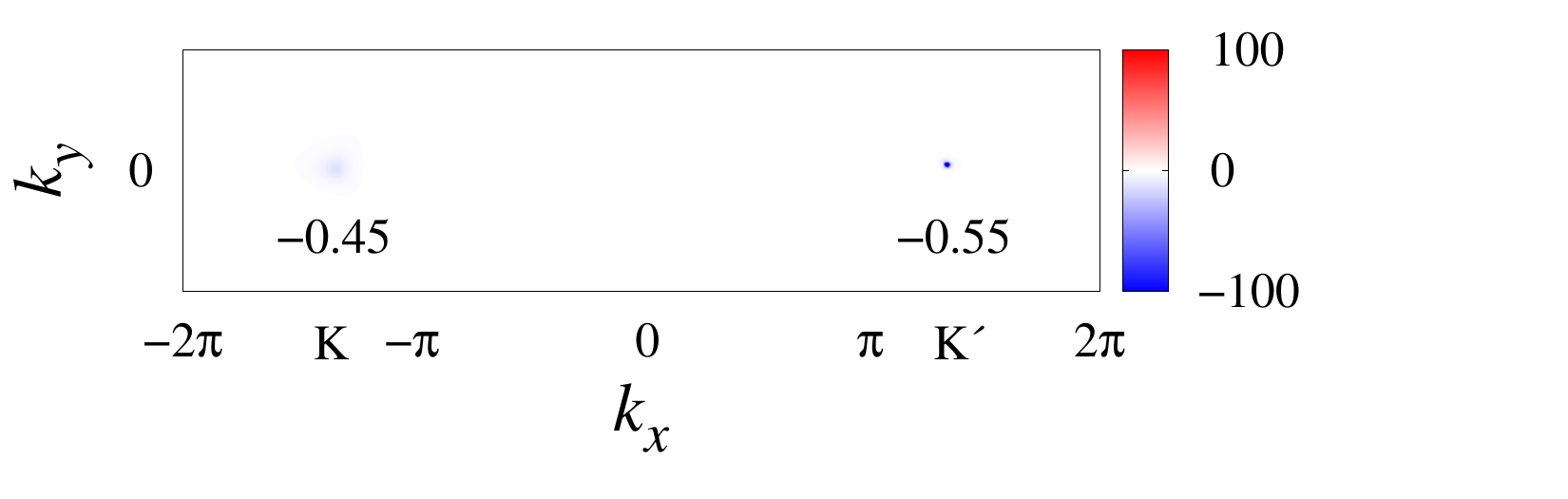}
	\includegraphics[width=8.5cm]{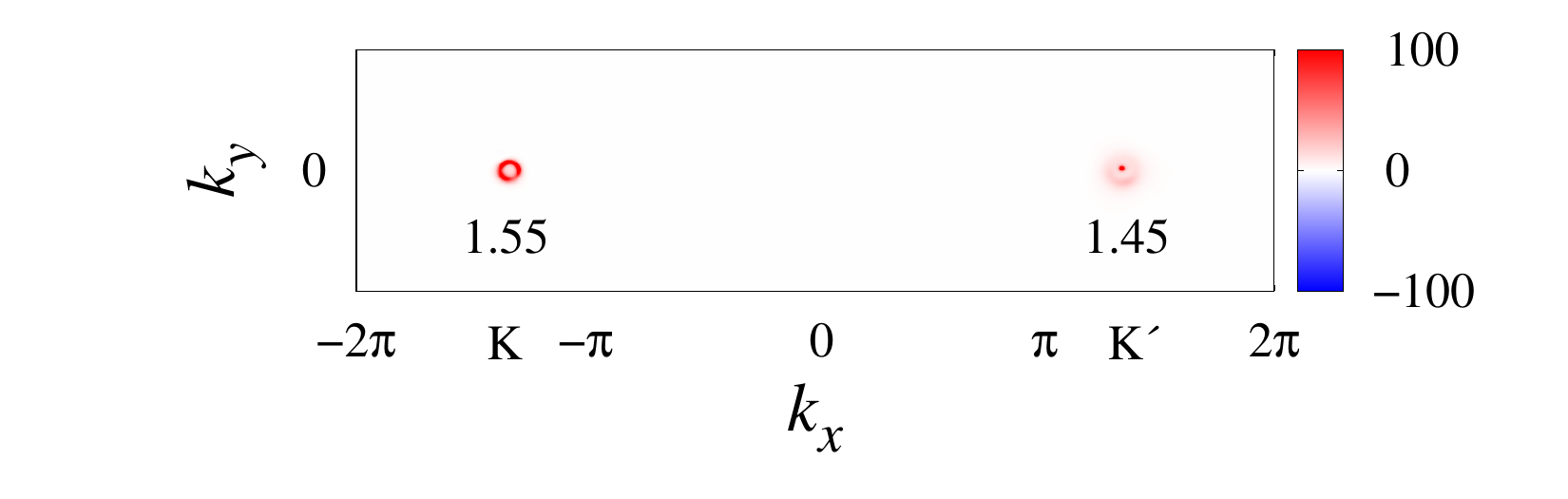}
	\includegraphics[width=8.5cm]{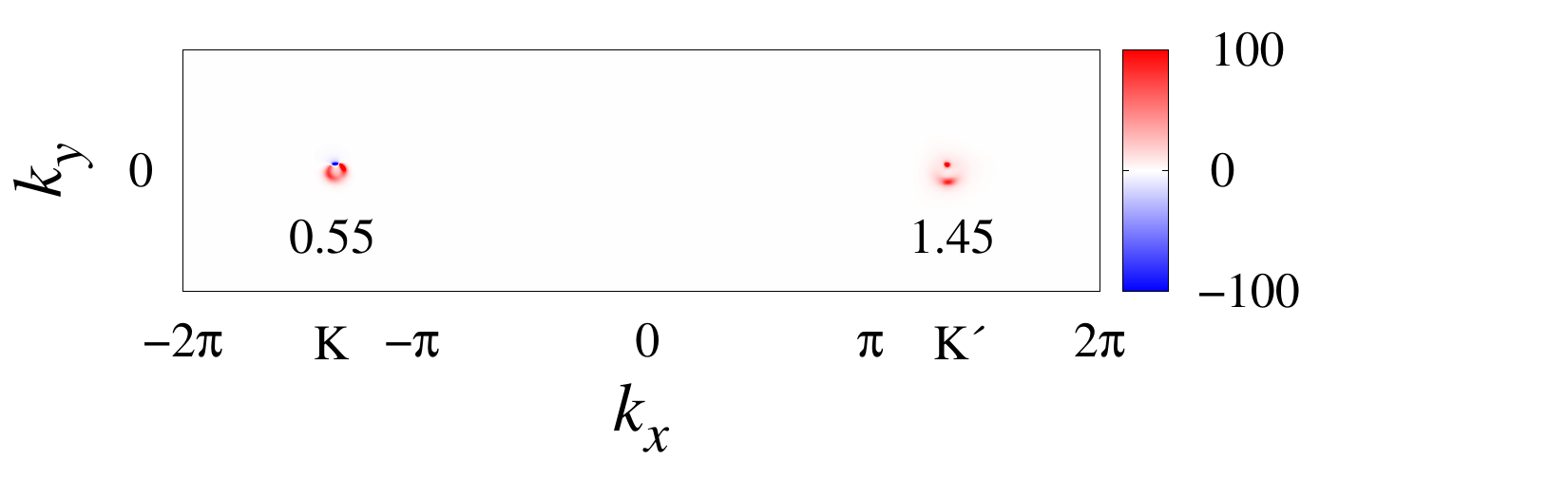}	
	\includegraphics[width=8.5cm]{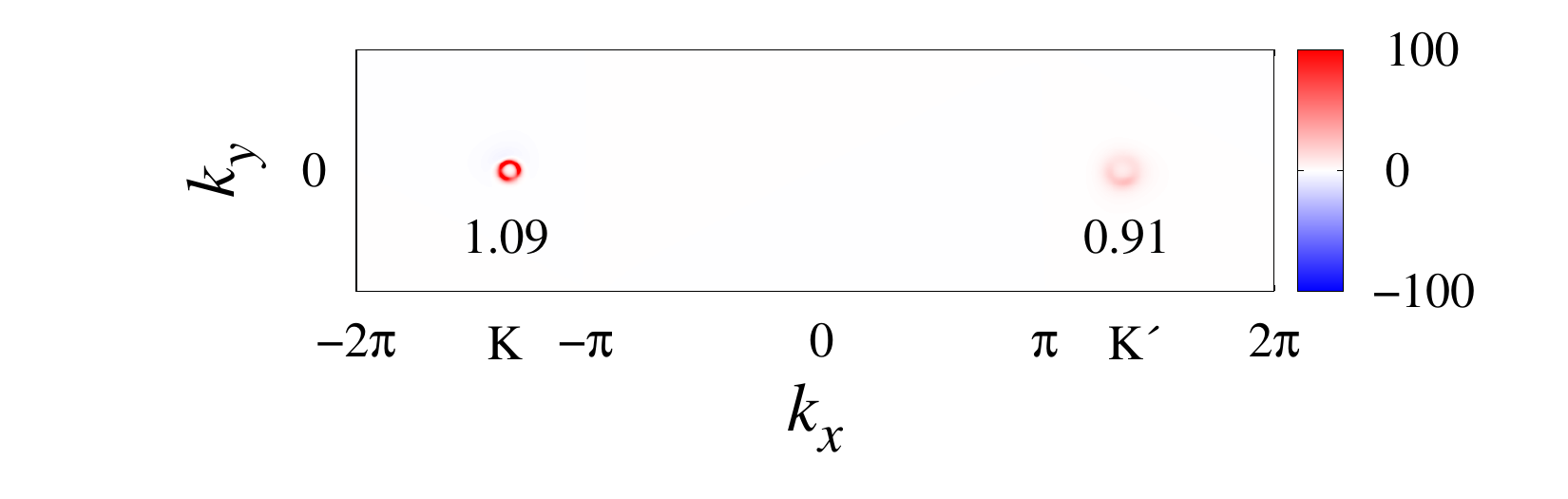}
	\includegraphics[width=8.5cm]{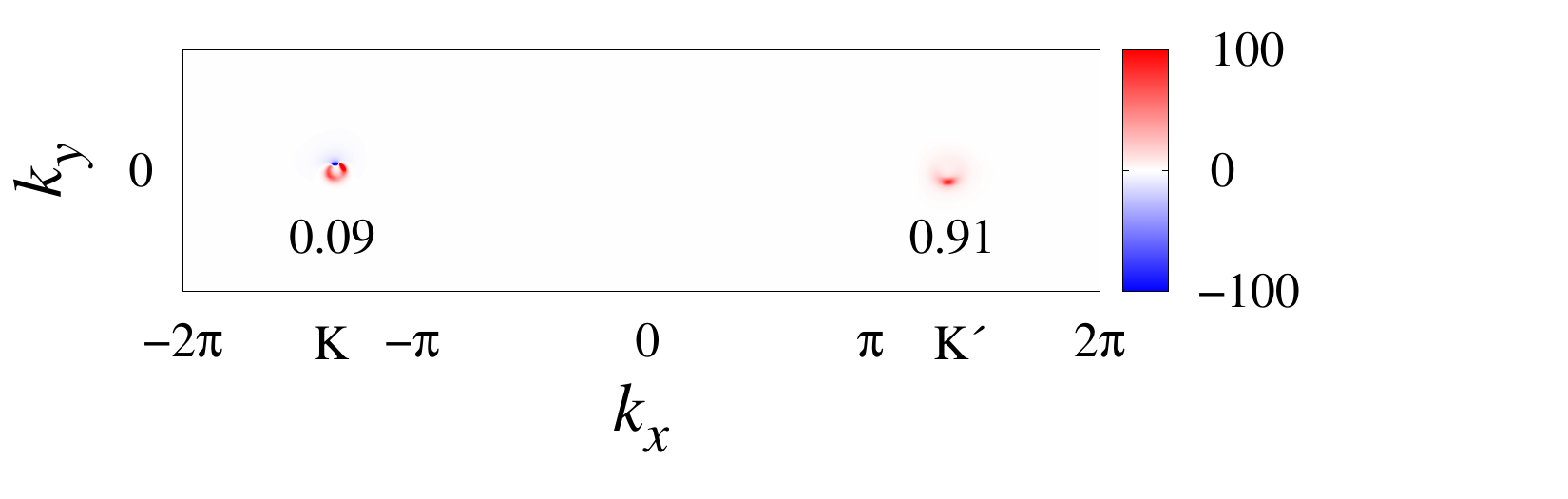}
	\caption{The distribution of Berry curvature in a Brillouin zone in the
			systems of $\theta_{T}=0.25$(left) and $\theta_{T}=0.7$(right). 
			The range of the Brillouin zone is $-2\pi<k_{x}<2\pi$ and
			$-\frac{1}{\sqrt{3}}<k_{y}<\frac{1}{\sqrt{3}}$. 
			The color on the
			momentum plane denotes Berry curvature ranging from -100 to 100. The
			numbers written in the plane are Chern numbers to which the Berry
			curvature is integrated over the half of the Brillouin zone including
			$K$ and $K^{\prime}$, respectively. 
			The first, second,
			and third rows correspond to the lower valence band, the 
			upper band valence band, and all the filled bands, respectively. 
}
			\label{BC} 
\end{figure*}

The energy dispersions at the transition points, plotted in
Fig.~\ref{tri_ED}(a) and (b), reveals the nature of two transitions.
At $\lambda_{E,c1}=\Delta$, the valence and the conduction band touches at
$K$ point and the Chern number decreases by one.
In contrast, at $\lambda_{E,c2}$ the energy dispersion exhibits three
band-touching points placed in the form of an equilateral triangle  
around  $K$ point, which increases the Chern number by three.
It is reminiscent of trigonal-warping deformation which is known to be induced
in graphene by Rashba spin-orbit interaction~\cite{Rakyta2010}.
The introduction of the sublattice potential shifts the topological 
transition point from
$\lambda_E=0$ to $\lambda_{E,c1}=\Delta$, and we presume that it gives rise to
additional fine splitting of the trigonal-warping deformation at $\lambda_{E,c2}$
from the $K$-point band-touching  at $\lambda_{E,c1}$.

For finite $\theta_T$, each of three band-touching points produces 
different phase boundary lines, as shown in Fig.~\ref{PD2}.
Two of them merge at finite $\theta_T$, forming a closed phase boundary line
which encloses a trivial phase with $C=0$.
Two typical energy dispersions on the closed phase boundary line are shown in
Figs.~\ref{tri_ED} (c) and (d).
They show a single band-touching point with a distorted trigonal-warping
deformation.
The phase boundary line generated by the third band-touching point  
is that separating $C=2$ phase from $C=1$ phase; 
this is the one shown in the global phase diagram of Fig.~\ref{pd}.
We can also observe that
metallic regions emerge around the region
where the closed phase boundary line is overlapped with that generated by 
the $K$-point band-touching for both systems.

We also display the band-touching momentum $\bm{k}_0$ on these phase boundaries
in Fig.~\ref{GPlocus2}.
As is discussed in the above, 
the three points at $\lambda_{E,c2}$ form an equilateral triangle, and 
two of them merge when $\theta_T$ is increased up to a critical value.
The third band-touching point goes towards $\Gamma$ point as $\lambda_E$ is
increased, and reaches close to $\Gamma$ point for very large $\lambda_E$.

\subsection{Berry curvatures and winding numbers }
 \begin{figure*}
 	\includegraphics[width=8cm]{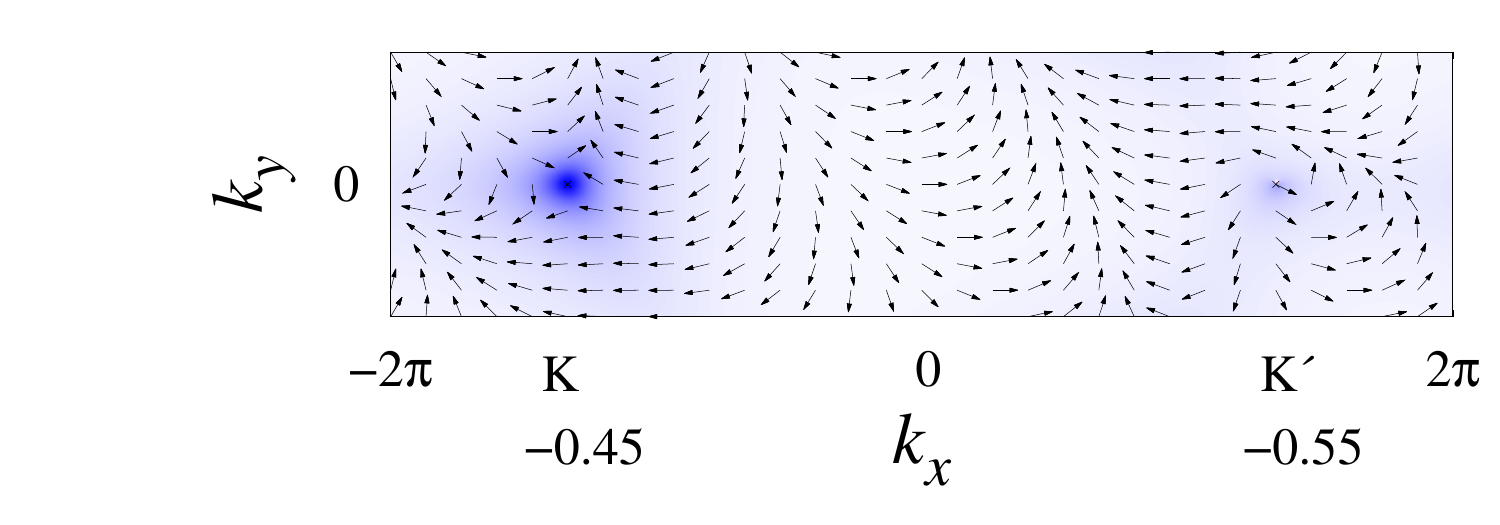}
 	\includegraphics[width=8cm]{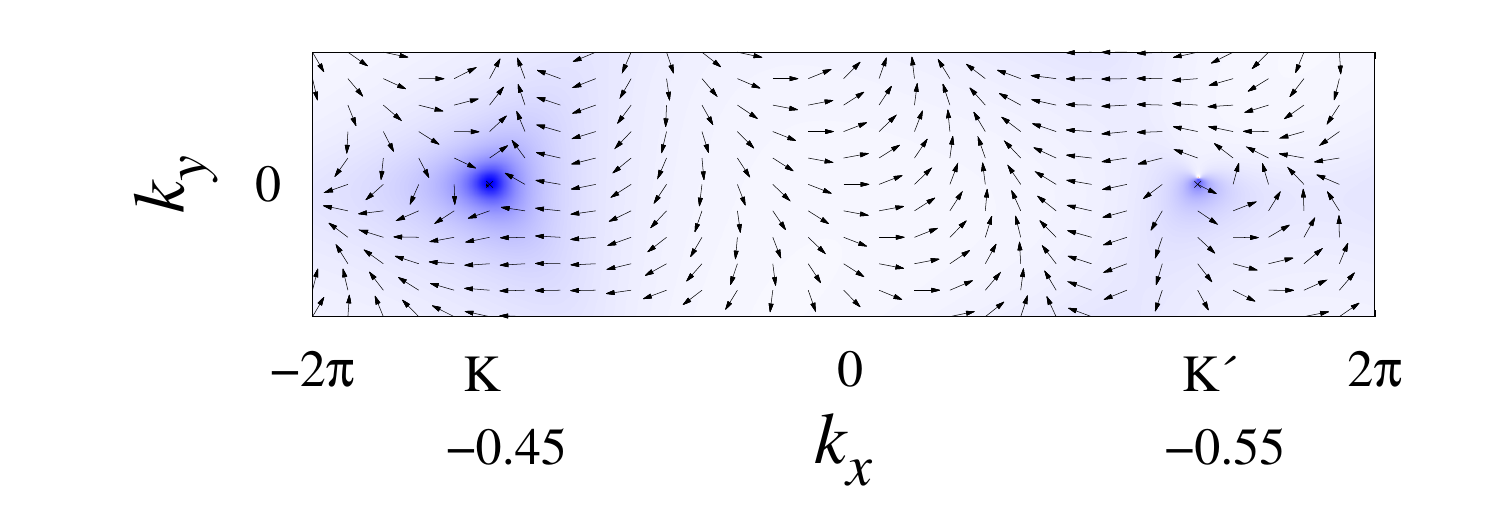}
 	\includegraphics[width=8cm]{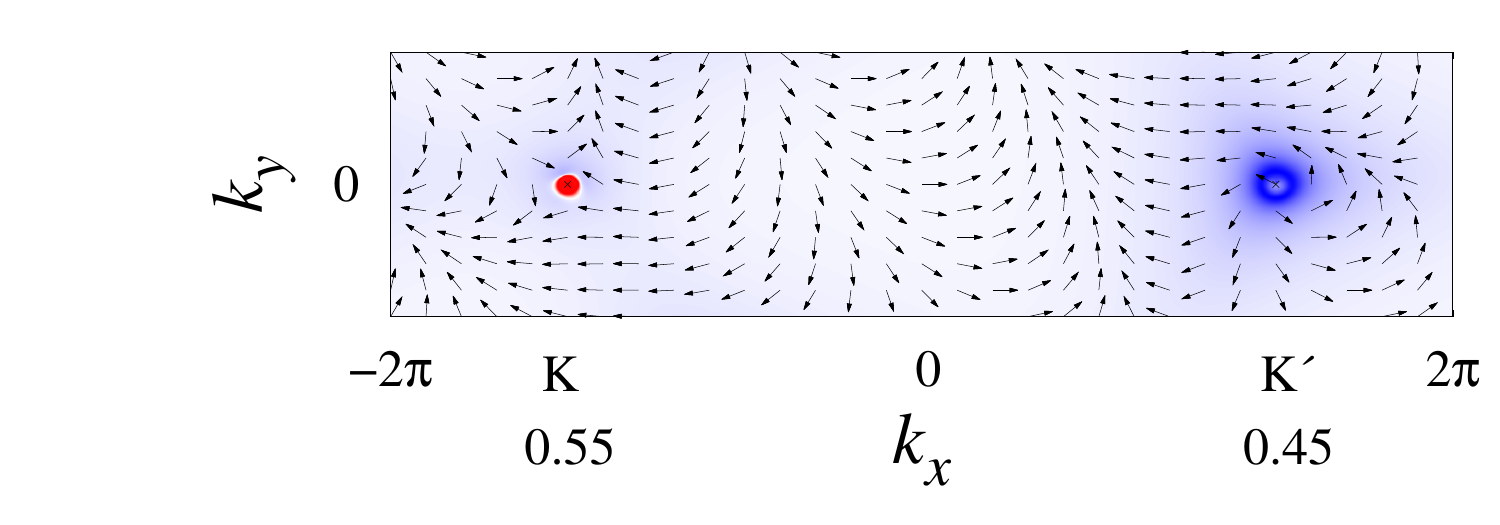}
 	\includegraphics[width=8cm]{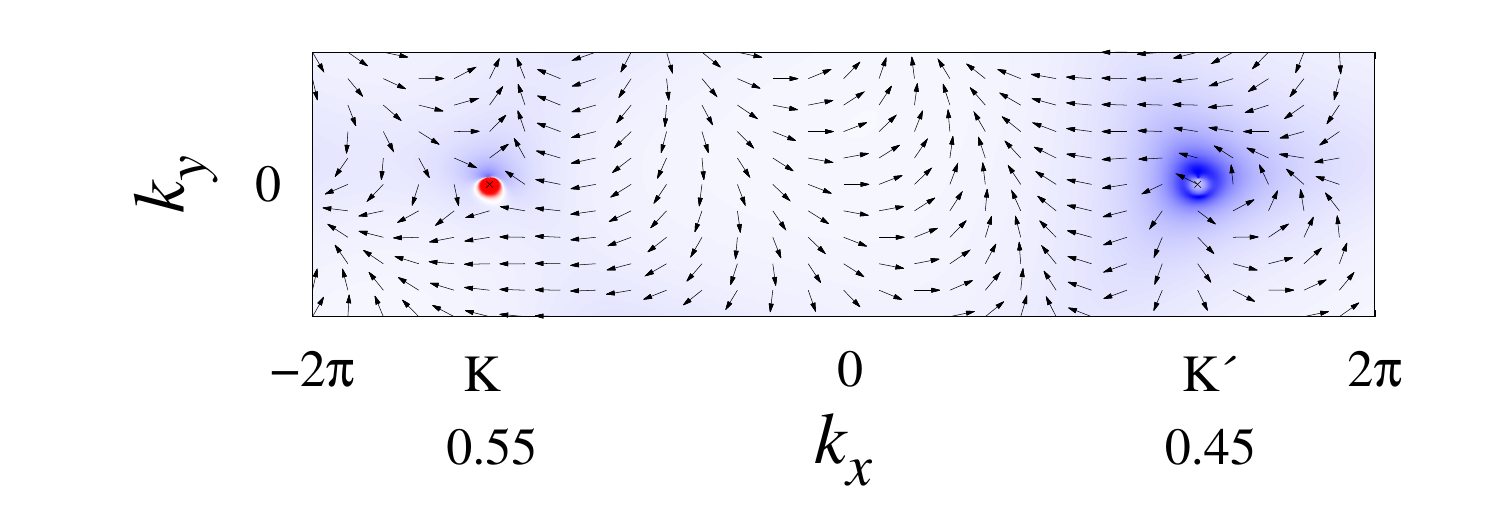}
 	\includegraphics[width=8cm]{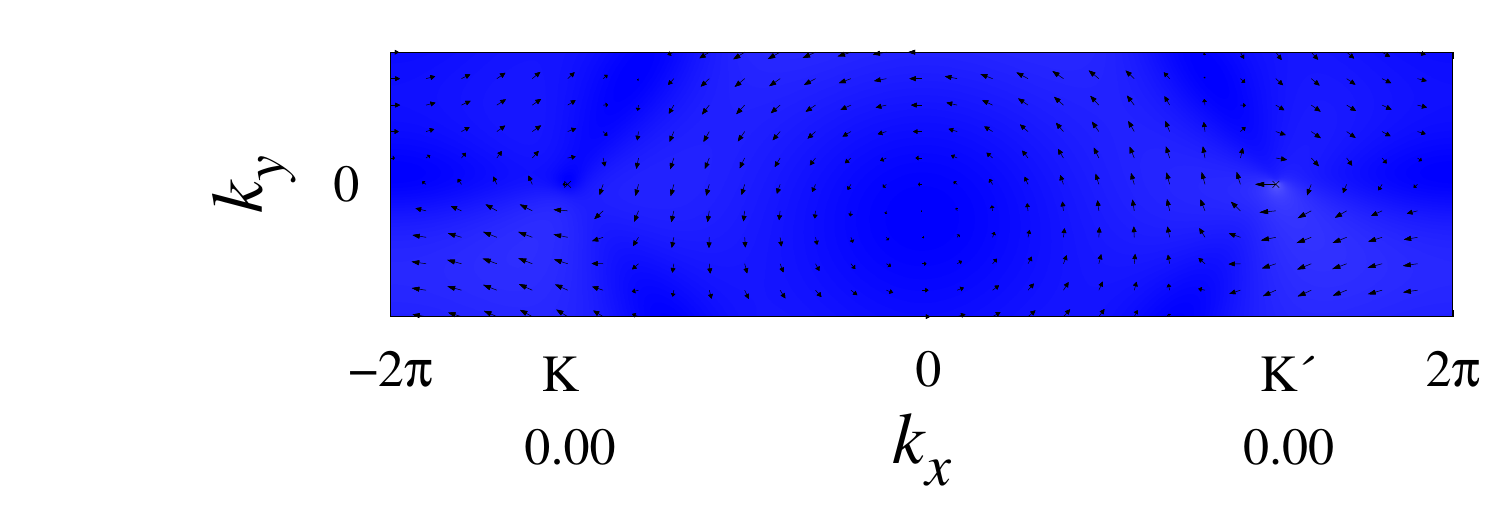}
 	\includegraphics[width=8cm]{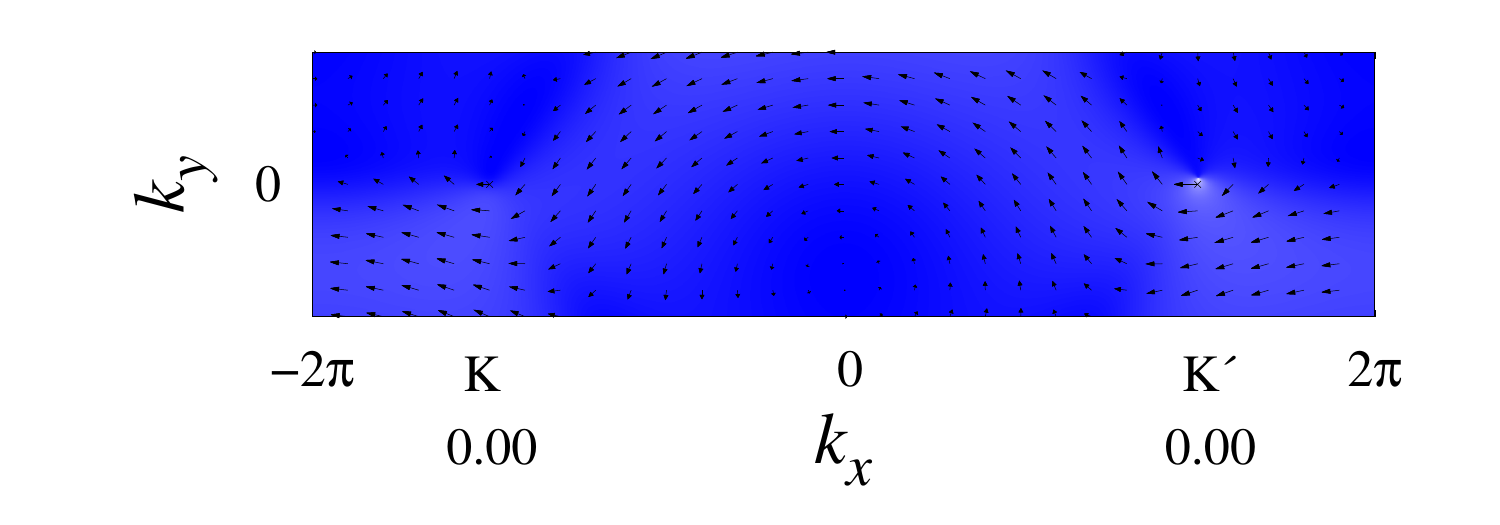}
 	\includegraphics[width=8cm]{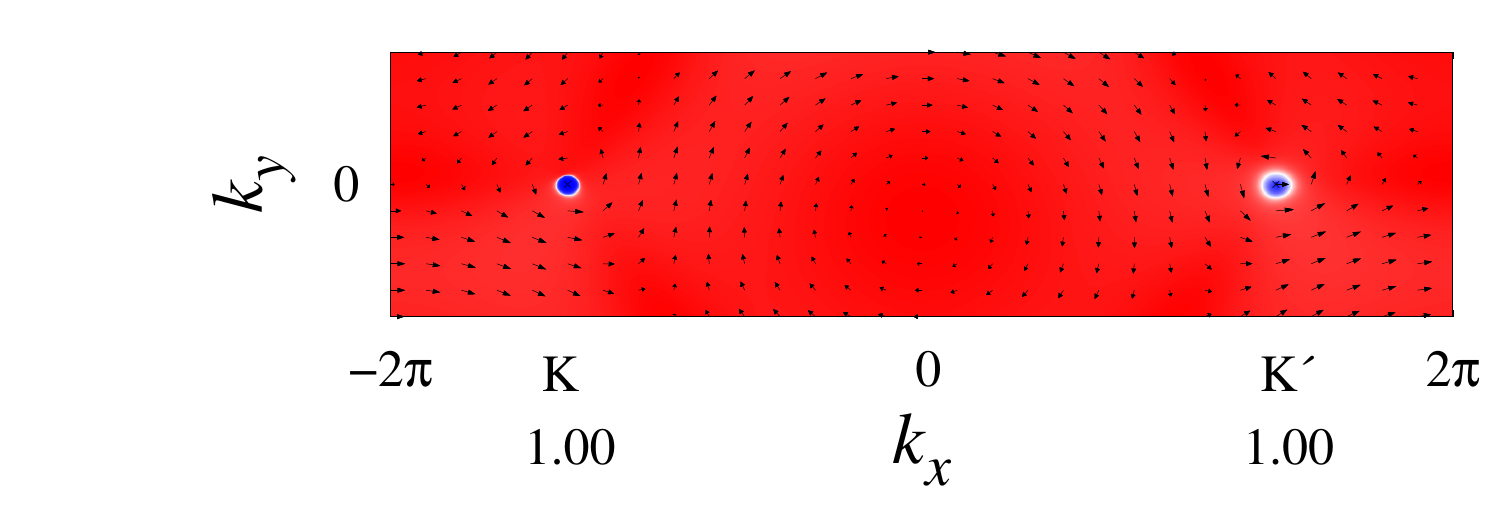}
 	\includegraphics[width=8cm]{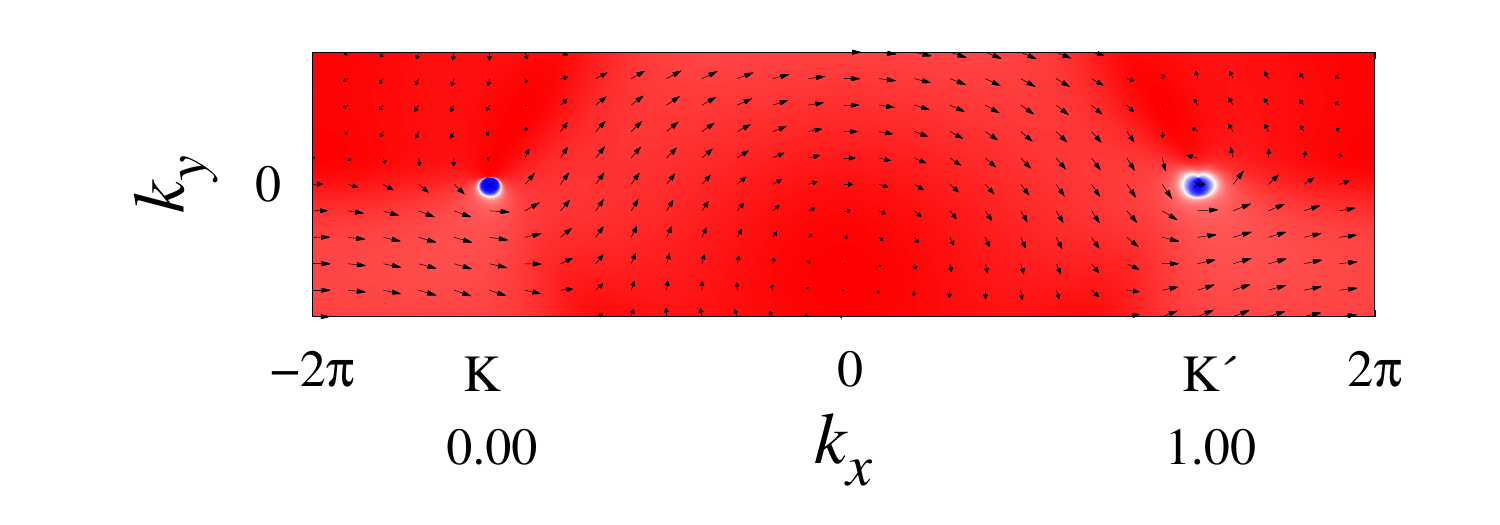}
 	\caption{Spin textures of pseudo and real spins for both valence
			bands in the cases of $\theta_{T}=0.25$(left) and
			$\theta_{T}=0.7$(right). 
			The upper(lower) two rows correspond to pseudo(real) spin textures. 
			The first and the third rows correspond to the lower valence band and
			the second and the fourth rows to the upper valence band.
			Arrows on the Brillouin zone
			represent the planar components of pseudo or real spins  and
			the color denotes those values of the $z$-component: blue$(-)$, white$(0)$, and red$(+)$. 
		The numbers written below $K$ or
			$K^{\prime}$ are the contribution to the winding number from
			the half of the Brillouin zones around the point.}\label{ST} 
 \end{figure*}

  In this section, we demonstrate topological properties of the nontrivial phases in terms of Berry
curvature and winding numbers.
We focus on two systems with $\theta_{T}=0.25$ and $\theta_{T}=0.7$ for $\lambda_{I}=-0.05$ and $\lambda_{E}=0.15$. 
The former and the latter systems exhibit the topological phases with $C=2$
and  $C=1$, respectively, as shown in Fig.~\ref{pd}(a).  

Figure~\ref{BC} shows the distribution of Berry curvature of the 
individual and all the valence bands.
In both valence bands, Berry curvature concentrates on $K$
and $K^{\prime}$. 
In the case of $\theta_{T}=0.25$, the Berry curvature in the
lower valence band contributes to the total Chern number negatively both in $K$
and $K^{\prime}$. 
However, those in the upper valence band are positive, which are much
larger than those in the lower valence band. 
As a result, both $K$ and
$K^{\prime}$ have positive Berry curvature peaks and their sum produces
Chern number  two.
In the case of $\theta_{T}=0.7$, the Berry curvature distributions are
more or less the same as in the case of $\theta_{T}=0.25$ except for the area
around $K$. 
An additional negative peak shows up as well as a positive peak near
$K$ in the upper valence band, and the Chern number of this area is 
reduced by one. 
Consequently, the total Chern number for $K$ is
less than that of $\theta_{T}=0.25$ by one. 
Thus, the phase transition from
$C=2$ to $C=1$ is attributed to the change of Berry curvature distribution
around $K$ in the upper valence band. 
 
 The topological properties can also be represented in terms of the winding 
		  numbers in spin textures~\cite{Qiao2012,
		  Fradkin2013,Bernevig2013,Fosel2017,Ren2016a,Nagaosa2013,Yu2010}. 
We calculate pseudo spin $\langle\bm{S}\rangle$ associated with two valleys and real spin
$\langle\bm{\sigma}\rangle$ in momentum space and 
the winding number $\omega$ in each texture is defined by
 \begin{equation}
 		\omega= - \frac{1}{4\pi}
			\int_{\rm BZ} \hat{\bm{M}}(\bm{k})
			\cdot\Big(\partial_{k_x}\hat{\bm{M}}(\bm{k})
			\times\partial_{k_y} \hat{\bm{M}}(\bm{k})
		\Big) d^2 k,
	\label{eq_wn_3d}
 \end{equation}
 where $\hat{\bm{M}}(\bm{k})$ is the unit vector in the direction of each spin
 at momentum $\bm{k}$. 
Chern number of the individual band is equal to the winding number, $C=\omega$~\cite{Fradkin2013}. 
 
 Figure~\ref{ST} describes the pseudo and real spin textures in both valence bands. 
In the case of $\theta_{T}=0.25$, merons and antimerons in pseudo spin textures
of lower and upper valence band cancel each other in both $K$ and $K^{\prime}$
while two skyrmions remain in both $K$ and $K^{\prime}$ in the real spin texture
of upper valence band. 
Thus, the nontrivial property of the system comes from
real spins in upper valence bands. 
As $\theta_{T}$ increases up to 0.7, 
the winding number changes from 1 to 0 due to the contribution in the vicinity of $K$ in the upper valence band. 
This is due to the destruction of a skyrmion at $K$ by the band-touching of the upper
valence and the lower conduction bands at $K$.

\section{Summary}
In summary, we have investigate the topological phase transition of the modified graphene
model under the twisted exchange couplings.
By the variation of the twist in the directions of two sublattice exchange
couplings we have successfully examined the nature of transitions between the
topological phases under uniform exchange couplings and those under staggered
exchange couplings.
The resulting phase diagrams have been found to exhibit rich phases.
We have performed the perturbative calculation in the twist angle, which was
successful in describing the phase transition line near the uniform exchange
couplings.
Topological objects in real and pseudo spin textures have been shown to the
source for the contribution to topological invariants of the system. 

Remarkably, we have discovered that
the transition from a trivial phase to a topological phase with Chern
number two in uniform exchange coupling is not a direct transition.
As the exchange coupling increases, the system first make a transition from 
the trivial phase to a topological
phase with Chern number reduced by one.
At a higher value of exchange coupling, the trigonal-warping deformation has
been found to drive the system to the topological phase with Chern number two.
The two close but separate transitions may have its origin in the interplay
by the Rashba spin-orbit coupling and the staggered sublattice potential.

\bibliography{QAHE}

\end{document}